\documentclass[reprint,amsmath,amssymb,aps]{revtex4-2}

\usepackage{graphicx}
\usepackage{dcolumn}
\usepackage{bm}
\usepackage{float}
\usepackage{amsmath}
\usepackage[utf8]{inputenc}
\usepackage{multirow}

\begin{document}

\title{Substrate disorder promotes cell motility in confluent tissues}

\author{Diogo E. P. Pinto$^{1,2}$, Margarida M. Telo da Gama$^{1,2}$ and Nuno A. M. Ara\'{u}jo$^{1,2}$}

\affiliation{$^{1}$ Centro de Física Teórica e Computacional, Faculdade de Ciências, Universidade de Lisboa, 1749-016 Lisboa, Portugal. \\ $^{2}$ Departamento de Física, Faculdade de Ciências, Universidade de Lisboa, 1749-016 Lisboa, Portugal.}

\begin{abstract}

\textit{In vivo} and \textit{in vitro} cells rely on the support of an underlying biocompatible substrate, such as the extracellular matrix or a culture substrate, to spread and proliferate. The mechanical and chemical properties of such structures play a central role in the dynamical and statistical properties of the tissue. At the cell scale, these substrates are highly disordered. Here, we investigate how spatial heterogeneities of the cell-substrate interaction influence the motility of the cells in a model confluent tissue. We use the Self-Propelled Voronoi model and describe the disorder as a spatially dependent preferred geometry of the individual cells. We found that when the characteristic length scale of the preferred geometry is smaller than the cell size, the tissue is less rigid than its homogeneous counterpart, with a consequent increase in cell motility. This result is in sharp contrast to what has been reported for tissues with heterogeneity in the mechanical properties of the individual cells, where the disorder favors rigidity. Using the fraction of rigid cells, we observe a collapse of the motility data for different model parameters and provide evidence that the rigidity transition in the model tissue is accompanied by the emergence of a spanning cluster of rigid cells.

\end{abstract}

\maketitle

\section{Introduction}

The idea of growing artificial cell tissues and organs has been around for several decades~\cite{Langer1993, Guimaraes2020}. This has spurred a truly multidisciplinary effort to understand the mechanisms responsible for the development of cell tissues and to search for novel strategies to tune the shape and mechanical properties of the tissue. Among those strategies is the use of biocompatible substrates~\cite{Langer1993, Iskratsch2014, Garreta2019}. An extensive body of research shows that the cell morphology and dynamics are sensitive to the physical and chemical properties of their underlying structure, be it the extracellular matrix or a culture substrate~\cite{Lo2000, Discher2005, Guo2006, Neuss2009, Tambe2011, Murrell2011, Song2013, Sunyer2016, Janmey2019}. For example, it has been shown that the substrate stiffness can significantly affect the geometry of cultured cells, including their spreading area~\cite{Yeung2005, Janmey2019}, volume~\cite{Guo2017}, and shape elongation~\cite{Devany2019}. Thus, irrespective of the biological effects, the physical interaction between the cells and their supporting structure plays a critical role in the mechanical properties of the tissue. This poses a great challenge due to the typical level of disorder involved~\citep{Kim2011, Miller2013, Hadden2017}.

Both \textit{in vivo} and \textit{in vitro}, the epithelial layer of cells is supported by a complex polymeric structure, the extracellular matrix (ECM), which constrains the collective behavior of the tissue~\cite{Wolf2013, Lo2000, Aznavoorian1990, Smith2006, Provenzano2008}. For example, it has been observed that cancerous cells alter the ECM in order to promote invasion through healthy tissue~\cite{Kraning-Rush2012, Levental2009}. The tumor microenvironment supports diverse mechanical and biochemical interactions during cancer progression, which plays a significant role in the degree of tumor malignancy and metastatic potential~\cite{Carmeliet2000, Egeblad2002}. Tumors act as local sources of ECM remodeling, resulting in heterogeneous spatial profiles of the ECM network~\cite{Wolf2009}. These profiles can then influence the migration of surrounding cells~\cite{Kraning-Rush2013}. By generating cell-scaled tracks along migratory paths, cells will not need to squeeze through or clear constrictive mechanical barriers~\cite{Kraning-Rush2013}. Thus, although the ECM is often quantified by bulk metrics, it has a high degree of heterogeneity, which in turn influences the tissue itself, in a way that is largely unclear.

Despite the broad range of physico-chemical processes, which in many cases are system dependent~\cite{Guimaraes2020}, there are convincing arguments that simple, mechanistic models can provide valuable insight into the dynamics of living systems~\cite{Fletcher2014, Farhadifar2007, Alert2020, Camley2017}. Several models have been proposed to understand their collective behavior, from single particle descriptions to density field models~\cite{Bi2014, Bi2015, Farhadifar2007, Fletcher2014, Camley2017, Kabla2012, Hernandez2021, Huang2021}. The Self-Propelled Voronoi model (SPV) has been one of the models of choice to study confluent tissues~\cite{Bi2016, Sussman2018a}. The degrees of freedom are the positions of the center of each cell and the cell shape is obtained by Voronoi tessellation~\cite{Kaliman2016}. The dynamics is governed by an energy functional that is quadratic in the area and perimeter of each (Voronoi) cell, thus making the interactions truly many body. The mechanical properties of the tissue are either solid or fluid like, depending on the strength of activity and shape parameter, $p_0$, of the individual cells~\cite{Bi2016}. The solid-like regime is characterized by a finite shear modulus, while in the fluid-like regime the shear modulus drops significantly~\cite{Bi2016, Sussman2018}. These results agree both quantitatively and qualitatively with experiments on monolayer tissues~\cite{Park2015, Grosser2021}.

Here, we investigate how heterogeneities on the substrate affect the mechanical properties of confluent tissues (see Fig.~\ref{fig1} a)). We describe the confluent tissue using the Self-Propelled Voronoi model, with a position-dependent shape parameter to account for spatial heterogeneities in the cell-substrate interaction. Previous works have established that cell shapes change as a function of substrate properties~\cite{Yeung2005} and in turn the cell shape governs the rate of cell diffusion in the tissue~\cite{Park2015}. Heterogeneity in the mechanical properties of the individual cells is known to substantially affect cell motility~\cite{Butcher2009, Sinkus2000}. For example, numerical simulations of the Vertex model suggest that, an heterogeneous distribution of the mechanical properties of individual cells favors rigidity and thus hinders cell motility~\cite{Li2019}. This particular type of cell disorder leads to larger tensions between adhered cells, which in turn gives rise to a percolating cluster of rigid cells responsible for the increase in the tissue rigidity. This result sheds light on the dynamics of cancer propagation, for cancer cells are usually softer than healthy ones~\cite{Ciasca2016, Alibert2017, Morawetz2017}. Here, we show that the opposite behavior is observed when the disorder is on the substrate (position dependent). For values of the characteristic length scale of the disorder lower than the typical cell size, the tissue is less rigid and cell motility is enhanced.

The paper is organized as follows. In section II we introduce the model. In section III we give an overview of the results. We consider the average diffusion coefficient of the cells to quantify the motility. In section IV we discuss the results obtained with a random substrate and discuss the collapse of the numerical data for different model parameters. In section V we focus on an averaged substrate and compare the results for three different disordered systems: cell, random and averaged substrates. In section VI we draw some conclusions and discuss practical implications.

\section{Model}

\begin{figure}[t]
	\includegraphics{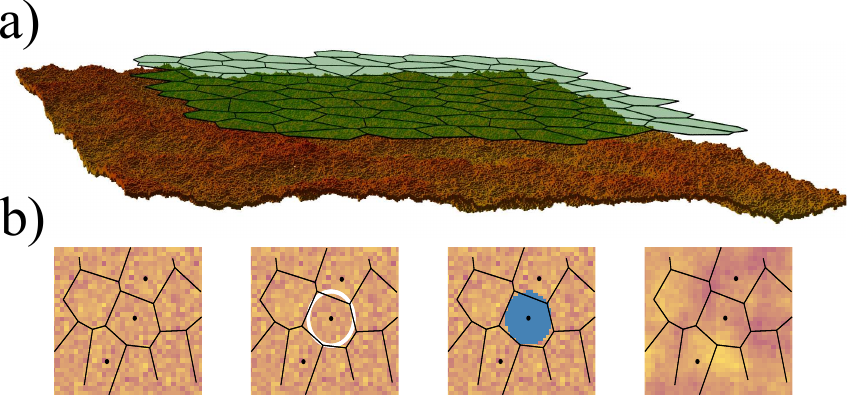}
	\caption{\label{fig1}a) Illustration of the model. We consider the 2D projection of the tissue (top, in green), described using the Self-Propelled Voronoi model, on a 2D heterogeneous substrate (bottom) where the value of the target shape index of the cells depends on the position. The color of the substrate is related to the value of the shape index in the square tiles, $p_{0, j}$, with red corresponding to higher values and yellow to lower ones. The height profile in the substrate is only meant to illustrate this heterogeneity. b) Schematic representation of the averaging process used on the random substrate. The averaging is performed by sweeping through each square tile and calculating the average of all points at a distance less than $\xi/2$. The white circle corresponds to the averaging radius of a given square tile close to a cell center. The blue shape corresponds to the square tiles used for the averaging. Here we have used approximately 80 blue square tiles, where each tile has a length two orders of magnitudes smaller than the typical length of a cell. The color of the substrate on the last panel represents the averaged substrate.}
\end{figure}

We model the confluent tissue as a monolayer of $N$ cells using the Self-Propelled Voronoi model~\cite{Bi2016}. Each cell $i$ is represented by its center $\textbf{r}_i$ and its shape is given by an instantaneous Voronoi tessellation of the space. The stochastic trajectory of each cell is obtained by solving a set of equations of motion in the overdamped regime,

\begin{equation} \label{motion}
	\frac{\mathrm{d}\textbf{r}_{i}}{\mathrm{d}t}=\mu \textbf{F}_{i}+v_{0}\hat{\textbf{n}}_{i} ,
\end{equation}

\noindent where $\textbf{F}_{i}$ is the net force acting on cell $i$, $\mu$ is the mobility of the cell, $v_{0}$ the self-propulsion speed, and $\hat{\textbf{n}}_{i}=(\cos\theta_{i},\sin\theta_{i})$ is a polarity vector which sets the direction of self-propulsion. For simplicity, we consider that $\theta_{i}$ is modeled by a stochastic process given by,

\begin{equation}
	\dot{\theta_{i}}=\eta_{i}(t), \hspace{0.1cm} \langle
	\eta_{i}(t)\eta_{j}(t')\rangle = 2D_{r}\delta(t-t')\delta_{ij} , 
\end{equation}

\noindent where $\eta_{i}(t)$ is an uncorrelated random process of zero mean and variance set by a rotational diffusion coefficient $D_{r}$.

The force $\textbf{F}_{i}$ describes the many-body cell-cell interaction and is given by $\textbf{F}_{i}=-\nabla E_i$, where $E_i$ is the energy functional for cell $i$~\cite{Farhadifar2007,Fletcher2014},

\begin{equation}
\label{eq::en.functional}
	E_{i}=K_{A}[A_{i}-A_{0}]^2+K_{P}[P_{i}-P_{0,i}]^2 \ \ ,
\end{equation}

\noindent where $A_{i}$ and $P_{i}$ are the area and perimeter of cell $i$, respectively, and $A_{0}$ and $P_{0,i}$ are their target values. The first term accounts for the cell incompressibility and the resistance to height fluctuations. The second term accounts for the active contractility of the actomyosin subcellular cortex and effective cell membrane tension, due to cell-cell adhesion and cortical tension. $K_A$ and $K_P$ are the area and perimeter moduli. By rescaling the energy in units of $K_AA_0^2$, we obtain four nondimensional quantities: two for the area and perimeter of the cell ($a_{i}=A_{i}/A_{0}$ and $p_{i}=P_{i}/\sqrt{A_{0}}$), a shape parameter $p_{0,i}=P_{0,i}/\sqrt{A_0}$, and the energy ratio $r=K_A A_0/K_p$ (see \textit{Supplemental Material} for details of the units system). Without loss of generality, in what follows, lengths are in units of $\sqrt{A_{0}}$ and time is in units of $1/(\mu K_{A}A_{0})$.

Cells diffuse with a diffusion coefficient $D$ that depends on four model parameters: the speed of self-propulsion $v_0$, the rotational diffusion $D_r$, the shape index $p_{0,i}$ of each cell $i$, and the energy ratio $r$. In the homogeneous case ($p_{0,i}\equiv ´p_{0}$ for all cells), for fixed values of $v_0$ and $D_r$, the model exhibits a rigidity transition at a value of the shape index $p_0=p_c$: from a fluid-like state, for $p_0>p_c$, where the cells diffuse on the substrate ($D\neq0$), to a solid-like state with finite shear modulus and negligible cell motility ($D\rightarrow 0$), for $p_0<p_c$~\cite{Bi2016,Merkel2019}. 

We consider a square substrate of length $L=\sqrt{N}$, where the value of the target shape index ($p_{0,i}$) is spatially dependent. The substrate is divided into square tiles with lattice constant $\delta$, in units of the cell diameter. Each square tile has a value of the target parameter randomly drawn from a Gaussian distribution with mean $\overline{p}_0$ and standard deviation $\sigma$. Throughout the dynamics, the value of the shape index, $p_{0,i}$, for each cell corresponds to the one in the underlying square tile. When a cell moves from one square tile to another, its target shape index (in Eq.~\eqref{eq::en.functional}) changes accordingly.

Since cells spread over an area larger than that of a square tile, we average the shape index over a distance on the random substrate to mimic, in a simplified way, the cells ability to probe its surroundings (see Fig.~\ref{fig1} b)). The averaged substrate has the same size as the random substrate and thus the same lattice constant $\delta$. The shape index on each square tile $j$, of the averaged substrate, is calculated by taking the average of the shape indices of the square tiles in the original substrate which have their centers at a distance smaller than $\xi/2$ from the centers of $j$. Thus, for a given $\xi$ the averaged substrate has mean $\overline{p}_0$ and standard deviation $\sigma/\sqrt{n}$, where $n$ is the number of square tiles inside the corresponding averaging circle. Some properties of the averaged substrate are further discussed in the \textit{Supplemental Material}.

To simulate the confluent tissue, we used a hybrid CPU/GPU software package, cellGPU~\citep{Sussman2017}, for the self-propelled Voronoi model. The equations of motion, Eq.~\eqref{motion}, are integrated using the Euler method, with a time step of $\Delta t=10^{-2}$. We impose periodic boundary conditions, $D_r=1$, $v_0=0.1$, and $r=1$. For the initial configuration, we generate $N$ positions at random and let the system relax over $10^4$ time steps. The random substrate consists of square tiles with lattice constant $\delta=0.03125$, in units of the cell diameter. This small value guarantees that there are no spatial correlations at the scale of the cell size and that cells are able to explore more than one substrate square tile even when they have very low motility. After the initial relaxation, the simulation is performed for another $10^6$ additional time steps.

\section{Overview}

To characterize the fluidity of the tissue, we measure the mean squared displacement from the initial position, averaging over all the cells ($\langle\Delta r^2(t)\rangle$) and we estimate the diffusion coefficient using:

\begin{equation}
\label{Diff}
D=\frac{\langle\Delta r^2(t)\rangle}{4t}, t\gg1 .
\end{equation}

This quantity is obtained numerically by running the simulations for $10^6$ time steps and calculating the slope of a linear fit, using the least squares method, of the mean squared displacement averaged over all the cells for all time steps above $10^5$. In the solid-like phase the cells are caged by their neighbors and few cell rearrangements occur. Thus, the mean squared displacement is characterized by an initial ballistic behavior ($\langle\Delta r^2\rangle\sim t^2$) but rapidly saturates. On the other hand, in the fluid-like phase, the cells are able to break free from their cages and the tissue flows. Thus, the mean squared displacement is diffusive ($\langle\Delta r^2\rangle\sim t$) asymptotically~\cite{Bi2016}. We have found that we can measure the diffusion coefficient reliably for time steps above $10^5$, in the fluid-like phase. As the solid-like phase is approached the fitting worsens, as the diffusion coefficient decreases to zero. Nonetheless, we still use the same technique.

\begin{figure}[t]
	\includegraphics{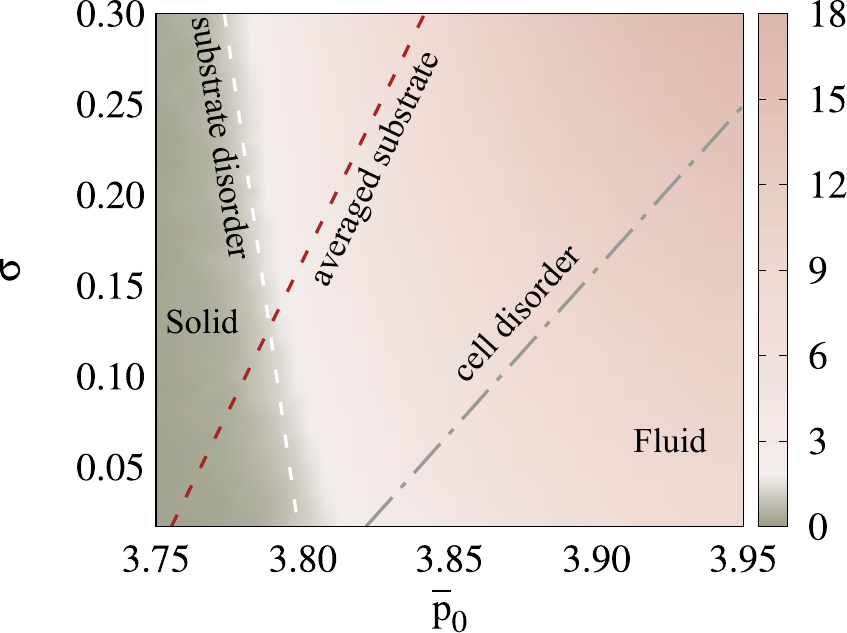}
	\caption{\label{fig2} Phase diagram of the tissue for substrate or cell disorder. On the vertical axis is the standard deviation of the Gaussian distribution, $\sigma$, and on the horizontal axis the mean, $\overline{p}_0$. The color gradient represents the diffusion coefficient of the tissue when using the random substrate disorder (given by Eq.~\eqref{Diff}), in units of $D^*\approx 9.04\times 10^{-5}$, which corresponds to the value of the diffusion coefficient at the onset of rigidity in the homogeneous system, i.~e.~, $p_0=3.8$. The white (dashed) line defines the threshold where the fraction of rigid cells (cells with $p_i<3.8$), forms a percolating cluster, $\sigma(\overline{p}_0)=-11.2\overline{p}_0+42.7$. Thus, it sets the onset of rigidity, in the presence of a disordered substrate. Results were obtained for $N=1024$ and averaged over $10$ samples. The gray (dot-dashed) line is obtained from Ref.~\cite{Li2019} for a tissue with heterogeneity in the mechanical properties of individual cells described by a cell-dependent shape index $p_{0,i}$, which is also drawn from a normal distribution with the same mean and standard deviation. In this case, the onset of rigidity is given by $\sigma(\overline{p}_0)=1.2\overline{p}_0-4.7$. The brown (dashed) line gives the onset of rigidity when an averaged substrate is used with correlation length of the order of the cell diameter. Here, the line is given by $\sigma(\overline{p}_0)=3.3\overline{p}_0-12.5$. This figure highlights the different effects of disorder. When the disorder is at the cell level the tissue becomes more rigid, while when it is spatially dependent (i.e., on the substrate) the tissue becomes less rigid when the substrate correlation length is less than the diameter of the cells, but more rigid when it is larger. The different phases are shown in the figure, where the tissue is marked solid or fluid. The lines do not meet at $\sigma=0$ since in Ref.~\cite{Li2019} the Vertex model was used rather than the Voronoi model used in this work.}
\end{figure}

A recent work studied the effect of heterogeneities in the mechanical properties of individual cells using the Vertex model~\cite{Li2019}. In this study, heterogeneity is introduced at the cell level by endowing each cell with a random shape index, $p_{0,i}$, chosen from a Gaussian distribution with mean $\overline{p}_0$ and standard deviation $\sigma$. The shape index of each cell is then constant over time. It was observed that the shear modulus increases with the disorder, $\sigma$, corresponding to a more rigid tissue. In what follows we compare the effect of the two types of disorder: substrate disorder, where the shape parameter is spatially dependent, and cell disorder, where the shape parameter is a time independent property of the cell as discussed in Ref.~\citep{Li2019}. For cell disorder, we have chosen the probability distribution of the shape index ($p_{0,i}$) to be Gaussian, parameterized by the same mean ($\overline{p}_0$) and standard deviation ($\sigma$). We note that both types of disorder are quenched as they do not evolve with the dynamics. Nevertheless, the substrate disorder is fixed in space, while that of the cells is carried by their motion in the fluid phase. In the rigid phase, both types of disorder are fixed in space, as cell motion practically ceases.

Figure~\ref{fig2} shows a diagram of the two-parameter space explored for a random substrate, where the color represents the average diffusion coefficient of the cells in the tissue using the random substrate disorder. We explored different values of the mean ($\overline{p}_0$) and standard deviation ($\sigma$) of the Gaussian distribution and observed that the diffusion coefficient increases for larger values of the disorder ($\sigma$), suggesting that the motility of the cells increases for substrates with larger gradients of the target shape index. We also show in the inset of Fig.~\ref{fig3} the increase of the diffusion coefficient with $\overline{p}_0$ and $\sigma$. This is in contrast with the cell disorder case where the rigidity of the tissue increases with increasing disorder~\cite{Li2019}. 

We plot three lines that give the onset of rigidity for the different types of disorder: cell disorder and substrate disorder with and without averaging (as described in the previous section). In depth details on how these lines are calculated are given in the following sections. We observe that the onset of rigidity in the tissue is accompanied by a percolation of rigid cells, defined as the cells with a perimeter smaller than a given shape index threshold, $p_i<\overline{p}_{0}^*$. When the substrate disorder is not averaged, the line is given by $\sigma(\overline{p}_0)=-11.2\overline{p}_0+42.7$, highlighting the increase in motility when the disorder ($\sigma$) increases. The cell disorder line is taken from Ref.~\cite{Li2019}, $\sigma(\overline{p}_0)=1.2\overline{p}_0-4.7$, and highlights the opposite behavior. Lastly, when the substrate with averaged disorder has a correlation length of the order of the cell diameter, $\xi=2$, the rigidity threshold is given by $\sigma(\overline{p}_0)=3.3\overline{p}_0-12.5$, which also exhibits an increase of tissue rigidity with disorder ($\sigma$). Thus, while for small correlation lengths the substrate disorder promotes cell motility, for correlation lengths of the order of the cell diameter, it gives rise to tissues with increased rigidity. Furthermore, the results for the averaged substrate approach those of cell disorder as the correlation length increases, as seen in the slopes of the two lines, suggesting a close relation between these types of disorder. We also note that the lines do not meet at $\sigma=0$ most likely because in Ref.~\cite{Li2019} a different model, i.e., the Vertex model was used.

\section{Random substrate}

First we focus on a random substrate with a lattice constant smaller than the cell size ($\delta=0.03125$). To calculate the lines separating the different regimes, we use a scaling \textit{Ansatz} to collapse the numerical data and estimate the transition between the solid and fluid-like regions. In Ref.~\cite{Li2019} a scaling \textit{Ansatz} is proposed for the shear modulus, which depends on the ratio between $\sigma$ and the distance to a threshold $|\overline{p}_0-\overline{p}_0^*|$, where $\overline{p}_0^*$ is the threshold value. The data was indeed observed to collapse. At $\overline{p}_0=\overline{p}_0^*$, $50\%$ of the cells are rigid with $p_{0,i}<\overline{p}_0^*$. This fraction decreases or increases above or below $\overline{p}_0^*$. This suggests that the fraction of rigid cells (cells with $p_{0,i}<\overline{p}_0^*$) plays an important role in driving the rigidity of the tissue. The fraction of rigid cells is defined as,

\begin{equation}
	\label{fr}
	f_r=\int_{-\infty}^{\overline{p}_0^{*}} \mathcal{F}_{\overline{p}_0, \sigma}(p_0)dp_0 = \frac{1}{2} \text{erfc}[(\overline{p}_0-\overline{p}_0^*)/\sqrt{2}\sigma],
\end{equation}

\noindent where $\sigma$ and $\overline{p}_0$ are the standard deviation and mean of the Gaussian distribution, respectively. For substrate disorder, the cells shape index changes frequently. Thus, the probability density function should also depend on the spatial distribution of the cells and we have no control over $\mathcal{F}$ in Eq.~\eqref{fr}. Nonetheless, for the parameters explored, we found that the cells have a distribution of the shape index, $p_{0,i}$, similar to the Gaussian used to generate the substrate disorder, with the same mean ($\overline{p}_0$) and standard deviation ($\sigma$). Thus, we use Eq.~\eqref{fr} to estimate the threshold of the transition, with $\overline{p}_0$ and $\sigma$ the same as those used for the substrate.

\begin{figure}[t]
	\includegraphics{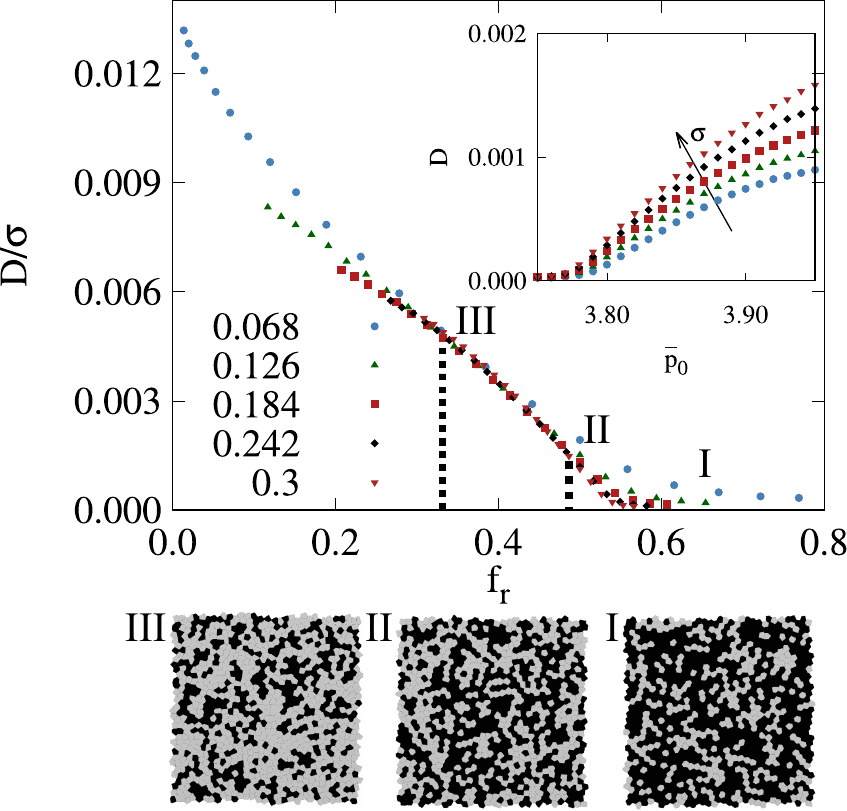}
	\caption{\label{fig3} Average cell diffusion coefficient as a function of the fraction of rigid cells, $f_r$. The diffusion coefficient is re-scaled by the standard deviation $\sigma$ to collapse the curves. These results were obtained for $N=1024$, $\sigma=0.068-0.3$ in steps of $0.058$, $\overline{p}_0=3.75-3.95$ in steps of $0.01$ and averaged over $100$ samples. The scaling suggests that the fraction of rigid cells drives the rigidity transition. In the inset are the individual curves to highlight the increase of the diffusion coefficient with the disorder ($\sigma$). Below the main plot are snapshots for different fractions of rigid cells ($f_r$), where rigid cells in black have a shape index below a given threshold ($p_{0,i}<\overline{p}_0^*$) and fluid cells in gray have, $p_{0,i}>\overline{p}_0^*$. We recall that $\overline{p}_0^*=3.8$ is the threshold for rigid cells.}
\end{figure}

Figure~\ref{fig3} illustrates the data collapse for the diffusion coefficient re-scaled by the standard deviation as a function of the fraction of rigid cells, $f_r$. To collapse the different curves we used $\overline{p}_0^*=3.80\pm0.01$ in Eq.~\eqref{fr}. We estimate this value numerically by using different threshold values ($\overline{p}_0^*$) and choosing the one for which we obtained the best data collapse. Figure~\ref{fig3} (bottom panels) illustrates the evolution of the rigid cluster, which corresponds to the largest cluster of rigid cells ($p_{0,i}<\overline{p}_0^*$) in black. We found that this cluster decreases as the standard deviation ($\sigma$) or the mean ($\overline{p}_0$) increase. The scaling \textit{Ansatz} suggests that the rigidity is driven by cells with a shape index, $p_{0,i}$, smaller than a threshold $\overline{p}_0^*$. As a result, we can estimate the onset of rigidity by measuring the threshold where the rigid cells form a system spanning cluster.

\begin{figure}[t]
	\includegraphics{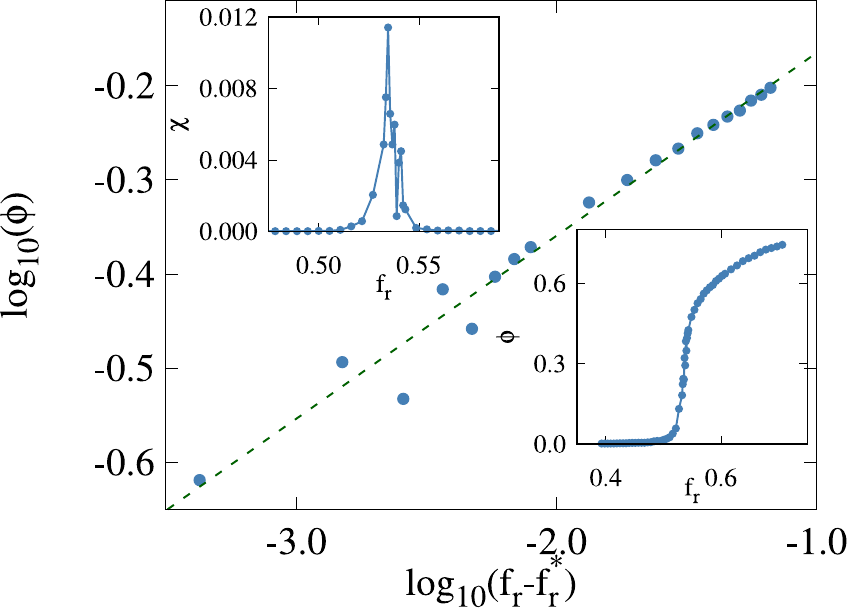}
	\caption{\label{fig4} Percolation at the rigidity transition. In the main plot is the fraction of rigid cells in the largest cluster, $\phi$, as a function of $(f_r-f_r^*)$, where $f_r^*\approx0.5353$, was calculated from the peak in the variance (top left inset). In the bottom right inset is $\phi$ as a function of the fraction of rigid cells, $f_r$. From the slope of the curve in the main plot we estimate the exponent $\beta\approx0.194\pm0.007$, which is consistent with that for 2D random percolation, $\beta=5/36$. The results were obtained for $N=16384$ and averaged over $10$ samples.}
\end{figure}

To analyze the percolation of rigid cells we measure the fraction $\phi$ of all rigid cells ($p_{0,i}<\overline{p}_0^*$) in the largest cluster. To estimate the percolation threshold, $f_r^*$, we consider the value at which the variance $\chi=\langle \phi^2 \rangle-\langle \phi \rangle^2$ is maximum. In the \textit{Supplemental Material} it is shown that $\chi$ has a peak around $f_r^*\approx0.484$ signaling the onset of percolation. When comparing to the data collapse in Fig.~\ref{fig3}, this threshold is not consistent with the point at which the diffusion coefficient starts increasing from small values ($D>10^{-4}$). From the numerical results, the distribution of perimeters of the cells, $p_i$, also follows a Gaussian distribution with mean $\overline{p}_0$ and standard deviation $\sigma$. Thus, we now consider a different criteria for rigid cells. We redefine a rigid cell as one with a perimeter ($p_i$) below the shape index threshold, $p_{i}<\overline{p}_0^*$. We considered the same threshold, $\overline{p}_0^*$, since the distribution of perimeters ($p_i$) and shape index ($p_{0,i}$) are similar. Thus, in our case, Eq.~\eqref{fr} is equivalent when using the distribution for either $p_i$ or $p_{0,i}$. We hypothesize that since $p_i$ is related to the tension in the cells ($\tau\sim p_i-p_{0,i}$) it can also be responsible for the decreased rigidity of the tissue, as detailed in previous works~\cite{Li2019, Bi2016}. We observe from the inset of Fig.~\ref{fig4} that the corresponding $\chi$ has a peak around $f_r^*\approx0.534$ signaling the onset of percolation, in line with the results from the data collapse in Fig.~\ref{fig3}. From a finite size scaling analysis, using the shift of the peak in the variance ($\chi$) with $N$, we estimate the threshold $f_r^*\approx 0.5353\pm 0.0003$. Using this value, we calculate the scaling exponent (shown in the main plot), $\beta\approx0.194\pm0.007$. Larger simulations are needed to estimate the exponents with higher precision, which is beyond the scope of this work. Nonetheless, the obtained value of $\beta$, is consistent with that for random percolation, $\beta=5/36$~\citep{Stauffer2018, Sahini1994}.

In Fig.~\ref{fig2} we plot in white the line corresponding to $f_r^*\approx0.5353$ which sets the percolation threshold for the disordered substrate. The gray line corresponds to $f_r^*\approx0.21$ taken from Ref.~\cite{Li2019} as the onset of rigidity in the cell disordered system. This highlights the differences between the two types of disorder and how they change the mechanical properties of the tissue. In the cell disordered case, the heterogeneity increases the tensions, $\tau\sim p_i-p_{0,i}$, leading to a more rigid tissue. For the substrate disorder, larger tensions can be observed at higher values of the disorder ($\sigma$) but the average tension in the tissue decreases. Furthermore the distribution of tensions becomes more symmetrical, which promotes the fluid-like state (see \textit{Supplemental Material}).

\section{Averaged substrate}

In a cell tissue, one expects that one cell senses a region of the substrate rather than a single point. In order to account for this effect, we consider next the averaged substrate described in the methods section. The averaging is aimed to mimic the process through which a cell senses a given area under it and thus responds. To each square tile $j$ in the averaged substrate corresponds a value of the shape index, $p_{0,j}$, which is the average of the shape index in the square tiles within a distance $\xi/2$ from $j$. The length scale $\xi$ sets the diameter of the circle used to calculate the averaged substrate disorder and thus sets the correlation length (as discussed in the \textit{Supplemental Material}). Figure~\ref{fig5} d) shows some examples of averaged substrates (bottom row) with the corresponding tissues (top row).

Figure \ref{fig5} a) depicts the diffusion coefficients measured for four different systems as a function of the mean, $\overline{p}_0$. We consider a random substrate (S, $\xi=0.03125$), a homogeneous tissue (H), a tissue with cell disorder (C) and an averaged substrate (S, $\xi=2$). Both substrate and cell disordered systems have a disorder dispersion $\sigma=0.184$. We confirm that, the cell disorder decreases the motility of the cells while the random substrate increases it. However, if the correlation length of the substrate is of the order of the typical cell diameter (or larger) then the cells become less mobile than in the homogeneous case. Thus, substrate disorder with large correlation lengths can also lead to more rigid tissues. In Fig.~\ref{fig5} b) this is shown for three different values of the standard deviation ($\sigma$). We find that while for correlation lengths lower than the typical cell diameter, $\xi<1$, more disordered substrates lead to larger diffusion coefficients than in the homogeneous case, for correlation lengths above the typical cell diameter, $\xi>1$, substrate disorder decreases cell diffusion. This happens since for large correlation lengths the cells adapt to the substrate smoothly as the gradient of the shape index, $p_{0,i}$, is small. By contrast, for smaller correlation lengths, $\xi$, the cells change their shape index $p_{0,i}$ quite rapidly (the dependence of the diffusion coefficient with $\xi$ is further explored in the \textit{Supplemental Material}). This leads to different behaviors of the tension distribution in the tissue. For $\xi<1$, as the dispersion, $\sigma$, increases, the variance of the tensions also increases while its average value decreases promoting fluid-like tissues. For $\xi>1$ the variance of the tensions is almost constant while their average value increases leading to more rigid tissues (see \textit{Supplemental Material}).

\begin{figure*}[t]
	\includegraphics{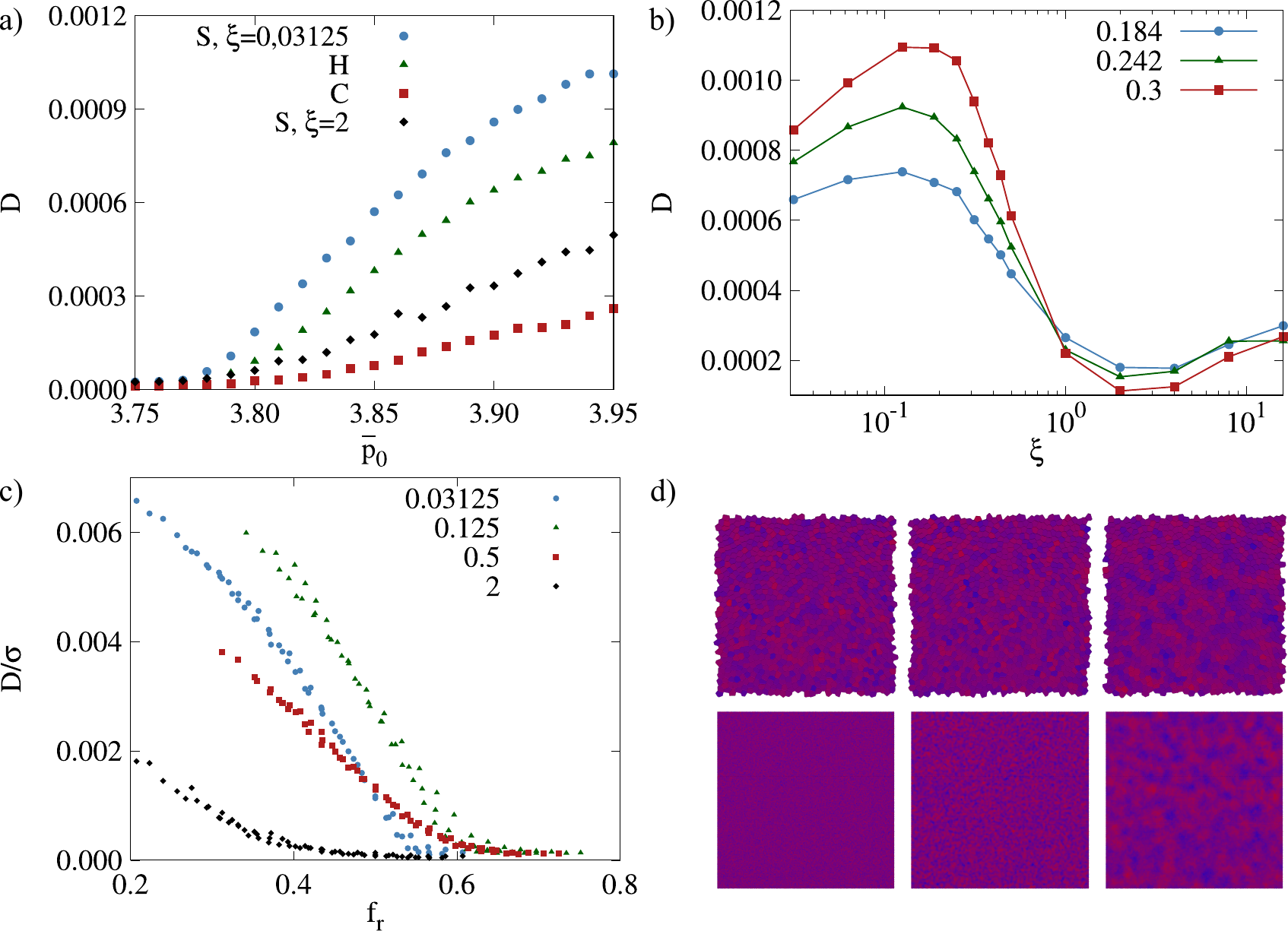}
	\caption{\label{fig5} Effect of substrate heterogeneities on cell motility. In a) the diffusion coefficients are plotted for different types of heterogeneity: ``S'' is for substrate disorder, not averaged for a correlation length $\xi=0.03125$ and averaged for $\xi=2$. ``C'' is for cell disorder, as in Ref.~\citep{Li2019}, where each cell has a random shape index $p_{0,i}$ from a normal distribution, which remains constant. ``H'' is for the homogeneous tissue. Panel b) illustrates how the diffusion coefficient varies with the substrate correlation length $\xi$, for a mean $\overline{p}_0=3.85$. In c) the diffusion coefficient re-scaled by the standard deviation ($\sigma$) is plotted as a function of the fraction of rigid cells, $f_r$, for four different correlation lengths, $\xi$. In d) are schematic representations of the tissue (top) and the substrate (bottom) for correlation lengths $\xi=0.125, 0.5, 2$ respectively. These results were obtained using $N=1024$ and averaged over $10$ different samples. We found that although the mechanical properties of the tissue change with the correlation length, $\xi$, the curves collapse with the fraction of rigid cells, $f_r$, suggesting that the percolation of rigid cells still drives the tissue rigidity. A correlation length of $\xi=1$, is found above which the response of the tissue to the substrate disorder changes, with higher disorder, $\sigma$, leading to a more rigid tissue.}
\end{figure*}

Both types of behavior, however, are related to the percolation argument developed above. In Fig.~\ref{fig5} c) we report data for different values of the correlation length ($\xi$) which is collapsed using the same scaling as in Fig.~\ref{fig3}. Thus, the behavior is driven by the percolation of rigid cells. As the correlation length ($\xi$) changes, the threshold values for the percolation transition also change ($\overline{p}_0^*(\xi)$ and $f_r^*(\xi)$). We note that for $\xi=2$, where the tissue is the most rigid, the percolation threshold for the fraction of rigid cells, $f_r^*\approx 0.38$, tends to the value reported in Ref.~\cite{Li2019}, which is consistent with a more rigid tissue than for the homogeneous substrate.

\section{Conclusion}

We studied the effects of spatial disorder of the cell-substrate interaction on the motility of the cells in a confluent tissue. We used the self-propelled Voronoi model, where the preferred geometry of each cell, $p_{0,i}$, depends on its spatial position. To model the spatial heterogeneities we divided the surface of the substrate into square tiles, where each tile has a value of the shape index, $p_{0,j}$, drawn from a Gaussian distribution with mean $\overline{p}_0$ and standard deviation $\sigma$. We also considered a more realistic description of an averaged substrate, where cells respond to a local averaged disorder. We introduced a correlation length for such an average and showed that for correlation lengths smaller than the cell diameter, the motility of the cells increases. For correlation lengths larger than the cell diameter, the disorder makes the tissue more rigid, by decreasing the cell motility. This is in contrast to what is known for tissues with disorder in the mechanical properties of the cells. For those tissues, the rigidity increases with the level of disorder~\citep{Li2019}. Our results suggest that, for smaller correlation lengths, the random change in the shape index leads to a more symmetrical tension distribution with lower average values, characteristic of more motile cells. For larger correlation lengths, the cells will have more time to adapt and the distribution of tensions shifts towards larger values characteristic of a more rigid tissue. We also note that for the largest values of the correlation length where the tissue is most solid like, our results approach those of the cell disorder reported in Ref.~\cite{Li2019}. This suggests that these two types of disorder are closely related.

We also show that our results for a given correlation length may be collapsed onto a single curve if we use the fraction of rigid cells (Eq.~\eqref{fr}) as the control parameter. This suggests that the changes to the mechanics of the tissue are a consequence of the percolation of rigid cells, characterized by a perimeter smaller than a given threshold  $p_{i}<\overline{p}_0^*$. Using the fraction of rigid cells in the largest cluster, we obtained the threshold for the reported increase in motility, $f_r^*\approx0.5353\pm0.0003$, for a completely random substrate without averaging, and a scaling exponent $\beta\approx0.194\pm0.007$. For larger values of the correlation length, our results suggest that the value of the threshold changes. Due to the symmetries of the model and the short-range nature of the correlations in the spatial distribution of the disorder, we hypothesize that the percolation transition belongs to the 2D random percolation universality class as corroborated by the value of the obtained exponents.

Although we focused on changes to the shape of the cells, more specifically the cells perimeter, it is expected that the substrate can affect other properties. Previous works~\citep{Sahu2020} have shown that the dynamics of mixtures is much less sensitive to differences in cell area than perimeter. Thus, we expect that, changes in area will lead to more subtle and potentially less marked effects.

These results can play an important role not only in tissue engineering, where the mechanical properties of the tissue are important~\citep{Guimaraes2020}, but also in the study of cancer where cells change the surrounding ECM in order to enhance their motility. Our results suggest that the underlying structure supporting the tissue, either the ECM or a culture substrate, should not be described using generalized bulk metrics since heterogenieties can play a relevant role in the tissue mechanics. Although we used a simplistic approach, these results should be robust to different substrate geometries. This also extends to curved substrates as long as the curvature does not play a major role in the tissue rigidity, as reported in previous studies~\citep{Sussman2020}.

Recently, it was shown that the cell adaptation time sets a minimum scale for the differences on the substrate that a cell can probe~\cite{Pinto2020}. Experimental values for the characteristic length of the changes in the substrate were used. Our results allow us to refine those calculations. The cell size sets the relevant length scale for adaptation, thus the typical size of epithelial cells in confluent tissues ($L\approx 20 \mu m$) sets the minimum length scale of the pattern for cells to be able to adapt. We can then calculate the expected adaptation time using the theory developed in Ref.~\cite{Pinto2020}. Using a typical diffusion coefficient for tissue cells of the order of $D\approx 0.1\mu m^2 min^{-1}$, we estimate that the typical time for a cell to adapt in a tissue interacting with a heterogeneous substrate is of the order of $\tau\approx 16$ hours. We have no knowledge of such measurements for cells in confluent tissues, but they fall within the relevant ranges for single cells adapting to heterogeneous substrates~\citep{Ebara2015}.

Here, we focused on a 2D description, but a 3D generalization is possible. In a simple generalization of the model to 3D~\cite{Merkel2018}, we expect similar results since a fluid to solid transition is present and the cells are able to diffuse throughout the tissue. If we consider a more realistic description of a 3D epithelial monolayer, then we would need a Vertex model along the lines of Ref.~\cite{Okuda2020} and characterize the apical and basal sides of the cells differently. Then, it is expected that the competition between the basal and apical perimeter difference plays a role in the diffusion of the cells. This would be interesting to explore in future studies.

We have also neglected both cell death and division. Due to modeling constraints, it is required that the number of cells remains constant throughout the simulation. Other works explored the effect of cellular division~\cite{Czajkowski2019}, and in the context of disordered media, it would be interesting to focus on how cell division or death play a role in tissue cell motility.

\section{Acknowledgments}

The authors acknowledge financial support from the Portuguese Foundation for Science and Technology (FCT) under Contracts no.~PTDC/FIS-MAC/28146/2017 (LISBOA-01-0145-FEDER-028146), UIDB/00618/2020, UIDP/00618/2020 and SFRH/BD/131158/2017.

\bibliography{Cells}

\end{document}